%Paper: chao-dyn/9406001
%From: yuliy@aphrodite.mathematik.Uni-Osnabrueck.DE (Yulik)
%Date: Thu, 9 Jun 1994 09:55:54 +0200
%Date (revised): Tue, 8 Nov 1994 18:38:38 +0100

%%%%%%%%%%%%%%%%%%%%%%%%%%%%%%%%%%%%%%%%%%%%%%%%%%%%%%%%%%%%%%%%%%%%%%%%
%%                                                                    %%
%%                        This is AMS-TeX                             %%
%%                                                                    %%
%%%%%%%%%%%%%%%%%%%%%%%%%%%%%%%%%%%%%%%%%%%%%%%%%%%%%%%%%%%%%%%%%%%%%%%%

\input amsppt.sty
\magnification=\magstep1

\define\Real{{\Bbb R}}\define\Rat{{\Bbb Q}}

\define\Int{{\Bbb Z}}

\def\qed{\hbox{${\vcenter{\vbox{
    \hrule height 0.4pt\hbox{\vrule width 0.4pt height 6pt
    \kern5pt\vrule width 0.4pt}\hrule height 0.4pt}}}$}}

\def\for{\  \hbox{ for } \ }

\def\and{\  \hbox{ and } \ }
\def\or{\  \hbox{ or } \ }

\def\Ga{\Gamma}

\def\s{\bold{s}}

\def\A{\Cal{A}}
\def\T{\Cal{T}}

\def\0{\bold{0}}

\title Complexity of Trajectories in Rectangular Billiards
\endtitle
\topmatter
\title Complexity of Trajectories in Rectangular Billiards
\endtitle
\author Yu. Baryshnikov \endauthor
\date May 1994 \enddate
\thanks The author was supported by DFG.
\newline  Department of Mathematics, University of Osnabr\"uck,
49069 Osnabr\"uck, Germany
\newline {\it E-mail address:} yuliy\@aphrodite.mathematik.uni-osnabrueck.de
\endthanks
\abstract
To a trajectory of a billiard in a parallelogram we assign its symbolic
trajectory --- the sequence of numbers of coordinate plane, to which
the faces met by the trajectory are parallel. The complexity of the
trajectory is the number of different words of length $n$ occurring in it.
We prove that for generic trajectories the complexity is well defined
and calculate it, confirming the conjecture of Arnoux, Mauduit, Shiokawa
and Tamura[AMST].
\endabstract
\endtopmatter

\document
\head 0. Introduction \endhead

Consider a rectangular billiard in $\Real ^{s+1}$, that is the
dynamical system
defined by the free motion of the point between
collisions with the boundary of the
billiard and elastic reflections at the collision instants such,
that the billiard is a
parallelogram with the faces parallel to the coordinate planes.

There is nothing especially
interesting in such a dynamical system as it is
equivalent to the trivial
system with the constant velocities on a torus.
However, the question of the coding
of the trajectory by means of
the listing of its consequent collisions with the boundary is meaningful and
have been attracting a lot of attention in the literature,
under different guises sometimes.

Specifically, one associates to a trajectory the infinite word in alphabet
$\A =\{ \0,\ldots,\s \}$ as follows: when the trajectory meets
a face of the parallelogram parallel to $j$-th coordinate plane, one
writes down $\bold j$. The resulting infinite word will be called
{\it symbolic trajectory}.
In exceptional cases a trajectory meets more than one face simultaneously,
but such cases are not generic and will not be considered.

The resulting symbolic trajectories
arise in numerous problems related to number theory,
quasicrystals, computer graphics etc. These
trajectories were abundantly studied in two-dimensional case
(where they bear also such names as Sturmian trajectories
or Beatty or Wythoff sequences); a sample bibliography can be found
in [B, LP, S]. Multidimensional generalizations are
investigated much less, though one can find quite
a lot of results on those in the mentioned papers.

The {\it complexity} of the symbolic trajectory is
defined as the number of
different words of the length
$n$ occurring in the associated symbolic trajectory considered as a
function of $n$. The problem of the
determination of the complexity of the rectangular
billiards was apparently first
studied by M. Morse and G.A.Hedlund in [MH], where it was
completely solved for
two-dimensional billiards.
They have shown, that the complexity is independent of
the trajectory (provided that the coordinate projection of velocity are
rationally incommensurable) and is equal to $n+1$.

Of course, the most striking fact here is the independence of the complexity
of particular trajectory. This independence persists in
three dimensional case, which was considered by Arnoux, Mauduit,
Shiokawa and Tamura [AMST]. They have shown, that
the complexity of symbolic trajectories is
$n^2+n+1$, as was conjectured by Rauzy
in [R3]. The authors made their own conjecture concerning
the complexity of the symbolic trajectories in multidimensional
case, basing on some quite mysterious assumption of symmetry
in $s$ and $n$. In fact, their formula follows immediately
from the independence of the trajectory result (see part 5
of the present paper).

Here we prove their conjecture, giving the general formula for
the complexity
of symbolic trajectories associated with rectangular
billiards in arbitrary dimension.

The method of the solution is as follows. The set of all subwords of
length $n$ of a symbolic trajectory
we call the
$n$-thesaurus.
First, we write down explicitly the
condition for a word of length $n$ to belong to $n$-thesaurus.
Further
we investigate the change of the $n$-thesaurus when
the vector of velocities varies. These changes happen
only when the inverse velocities
are rationally dependent, that is when a {\it resonance} occurs.
If the resonance is simple, that is there is at most one (up to multiples)
vanishing integer combination of inverse velocities, then we show that
there is a one-to-one correspondence between words leaving thesaurus and the
words coming into the thesaurus.  That proves,
that the complexity of the billiard is a well defined
function of $n$ only (and not of
the velocities). To finish, we calculate the
$n$-thesaurus for a special velocity
vector, which yields the main result of the paper:

\proclaim {Theorem}
If the inverse velocities of the free motion of the point are
independent over $\Rat$, then
the size of $n$-thesaurus is
$$
\sum_{k=0}^{\min (s,n)} {k !}{s \choose k} {n \choose k}
$$
\endproclaim

\head 1. Basic Constructions \endhead

Let $B \subset \Real^{s+1}$ be a rectangular area bounded by the hyperplanes
$\{x_i=0\}, \{x_i=l_i\}, i=1,\ldots,s$. Without loss of generality,
we will take all
$l_i=1$. This reduces the study of billiards in parallelograms
to that in the unit cube.
The movement of the particle in $B$ is defined as follows:
it moves freely with velocity $v=(v_0,\ldots,v_s)$ until
it reaches the boundary, where it reflects elastically
(that means that if the collision point belongs to the face
parallel to $i$-th coordinate plane, then $v_i \mapsto -v_i$).
The usual procedure of first
$2^{s+1}$-fold covering of $B$ by the torus and then
of the covering of the torus by
$\Real^{s+1}$ leads to the description of the
motion of the point as the projection of the
free motion of the point in $\Real^{s+1}$
with velocity $v$; the collision instants
correspond to the instants of the intersection
by the lifted trajectory of the
hyperplanes $x_j=n, n\in \Int$.
Excluding lower dimensional cases we can assume
that all $v_j \neq 0$; without loss of
generality, we can even take all $v_i >0$. For any $j$ the
instants of intersections of trajectories
with the hyperplanes $x_j=n$ form, obviously, an arithmetic
progression with the difference $a_j=(v_j)^{-1}$.
Thus the symbolic trajectory,
corresponding to a billiard trajectory with
the given velocity $v$ can be described as follows:
we mark points in $\Real$ belonging to $j$-th progression
by ${\bold j}$ and
then read all the marks in their natural order.
We assume that no point is marked
simultaneously by more than one numbers; this is true
for almost all trajectories with given
velocity vector (trajectories with the given velocity are parameterized
by their starting point modulo
shift along the trajectory,
which gives the $s$-dimensional torus as the space of trajectories;
trajectories for which the corresponding arithmetic progression
have common points form a countable union of $(s-1)$-dimensional
tori).
Trajectories for which no point is marked by more than one letter
(or, equivalently, which never hits $(s-1)$-dimensional faces of the cube)
will be called generic.

It is more convenient to work with
the vector $a=(a_0,\ldots,a_s)$ of {\it inverse velocities}
or differences of the arithmetic
progressions in question. We will say that a word $q$ in alphabet
$\A =\{ \0,\ldots,\s \}$ of length $n$ is $a$-admissible,
if there exists a generic trajectory
with velocities inverse to $a$,
such that $q$ is  a subword of the length $n$ in its symbolic
trajectory (it follows immediately, that in this case if
the differences $a_j$'s are $\Rat$-independent, then $q$ is a subword of
the symbolic trajectory for any generic trajectory
as the system is minimal).
The union of all $a$-admissible words of length $n$ is called
$n$-thesaurus for $a$ and is denoted as $\T(a)$.

We will represent the presence of $q$ in $\T(a)$
as some condition on a
polyhedron depending on the word and
velocities.
Introduce the following
$(3(s+1)+n)$-dimensional space $W$ with coordinates
$$
\eqalign{
x^-_0,\ldots,x^-_s;
x_1,\ldots &,x_n;
x^+_0,\ldots,x^+_s;\cr
a_0,\ldots &,a_s.
}
\tag1.1
$$

The meaning of the coordinates $x$ is the following. To an $n$-subword
of the symbolic trajectory $n$ consequent
instances correspond when the particle
hits the boundary. The numbers $x_1,\ldots, x_n$ represent just
these instants. The number $x^-_j$ ($x^+_j$, respectively) represents
the last moment before
$x_1$ ( after $x_n$) when the particle hits a
face parallel to $j$-th coordinate plane.

We will often consider $W$ as the direct sum of its $2(s+1)+n$-dimensional
$x$-part $W_x$ and $(s+1)$-dimensional $a$-part $W_a$;
the projection of $W$ on its $a$ part will be denoted
as $p_a$.

The conditions of precedence mentioned above are encoded by the following
inequalities defining a polyhedral cone $C \subset W$:
$$
\eqalign{
x^-_0 & \leq x_1, \ldots , x^-_s \leq x_1;\cr
 x_1 & \leq x_2 \leq \ldots  \leq x_n;\cr
 x_n & \leq x^+_0, \ldots, x_n \leq x^+_s.\cr
}
\tag1.2
$$

Now for any word $q$ of length $n$ in the alphabet $\A$ we define the
linear space $W(q) \in W$ as follows: let
$I_j=\{i^j_1,\ldots,i^j_{\vert I_j \vert }
\subset \{1,\ldots,n\}$ be the
subsequence of indices in $\{1,\ldots,n\}$
for which $q_i = j$. Then the linear
subspace $W(q)$ is defined by the conditions that the sequences
$$
x^-_j, x_{i^j_1}, \ldots, x_{i^j_{\vert I_j \vert}}, x^+_j
$$
form arithmetical progressions with the
differences $a_j$ for $j=0,\ldots,s$.

A simple count shows that the dimension of
$W(q)$ is $(2s+2)$ independently of $q$:
one can vary the first terms of the arithmetical
progressions and their differences arbitrarily.

Further, we define the convex polyhedral cone
$$
P(q)=C \cap W(q),
\tag1.3
$$

and its intersection with the fibers of the projection $p_a$ of
$W$ on its $a$-part

$$
P(q,a)=P(q) \cap p^{-1}_a(a).
\tag1.4
$$

These polyhedra play in sequel quite a fundamental role
because of the following

\proclaim {Lemma 1.5}
The word $q$ belongs to the thesaurus $\T(a)$
exactly when the following equivalent conditions hold:

1. $P(q)$ has the maximal dimension $2(s+1)$ and a
point $(x,a)\in W$ is interior in $P(q)$;

2. $P(q,a)$ has the
maximal dimension $(s+1)$ in the fiber $p^{-1}_a(a).$
\endproclaim

\demo{Proof} Let the word $q$ be a part of the symbolic trajectory
associated to a generic trajectory
with the vector of inverse
velocities $a$. Let
$\ldots t_{-1}, t_0, t_1,\ldots$
be the instances of collisions .
Then, assuming that
the word starts, say, at first term, one can by setting
$x_i=t_i, i=1,\ldots,n$ and attaching to $x_j^-$ ($x_j^+$)
the last moment of appearance
of ${\bold j}$ before $t_1$
(first moment of occurrence of ${\bold j}$ after $t_n$) get
an interior vector of $P(q)$ as all the inequalities
defining $P(q)$ are in fact strict at the point.
That means that a small vicinity of the point $(x,a)$ (with
$x=(x^-_0,\ldots,x^-_s,x_1,\ldots,x_n,x^+_0,\ldots,x^+_s)$)
in  $W(q)$ belongs to $P(q)$,
and the projection of the vicinity to $W_a$ is open there, that proves 1.

The assertion 2 follows immediately from 1.

Assume 2. It implies that there exists a point $(x,a)\in P(q,a)$
such that all the inequalities defining $P(q)$ are strictly satisfied.
Having such a vector $x$ one
easily constructs a piece of trajectory with
differences $a$ which has the word $q$ as
a part of its symbolic trajectory.
Extending the trajectory to both sides (which can be
done unambiguously) and disturbing a little
the arithmetical progressions to avoid
multiple points in their union
--- that always can be done as the fact that the
considered point is interior in $P(q,a)$ means that they are different
and small distortion do not change their
order --- gives the desired generic trajectory. \enddemo

The following statement generalizes Corollary 4 of [LP]:

\proclaim{Corollary 1.6} If a word $q$ belongs to the thesaurus $\T(a)$,
then the reversed word $q^{in}$ also belongs to it.
\endproclaim

\demo{Proof} The mapping
$$
x^-_j \mapsto -x^+_j , x_i \mapsto -x_{n-i},x^+_j \mapsto -x^-_j
$$
takes a point in $P(q,a)$ into a point in $P(q^{in},a)$.
\enddemo

\head 2. Changes of Thesaurus, Flows in Graphs and Linear Programming.
\endhead

Now we are going to study the changes of the
thesaurus when $a$ varies in a way.
Our goal will be to establish
the constancy of the thesaurus size. To prove it
we join two arbitrary vectors of inverse
velocities $a_1, a_2$ by the segment $I$
and consider its preimage under $p_a$ in $W$.
The intersection of the preimage with
any of the cones $P(q)$ is a convex polyhedron $P_I(q)$ again,
and the
Lemma 1.5 implies that the changes of the thesaurus
occur exactly in those points of $I$
where some of the polyhedra $P(q,a)$ lose their full dimension.
Such points we will call {\it critical}.

The totality of all polyhedra $P(q)$ is finite, and
each of them has a finite
number of faces, so the set of the critical points in $I$
is finite. Choose one such point
$a_*$ and two neighboring points $a_i,
a_o$, between which
no further critical point besides $a_*$ occurs.
That implies that the thesaurus {\it before} $a_*$
is that at $a_i$
and the thesaurus {\it after} $a_*$ is
that in $a_o$. Choose a linear function $l$
on the $W_a$ such that $l(a_i) < l(a_o)$,
and lift it to the whole $W$.

\proclaim {Lemma 2.1} The word $q$ belongs to
$\T(a_i) - \T(a_o)$ if and only if $P_I(q)$ has maximal
dimension $s+2$ and $l$ reaches its maximal value over
$P_I(q)$ on $P(q,a_*)$ --- a face of $P(q)_I$.
\endproclaim

\demo{Proof} The fact that $q \in \T(a_i)$ means by Lemma 1.5 that
$P(q,a)$ has the dimension $s+1$ for all $a \in I$ between
$a_i$ and $a_*$, thus the dimension of $P_I(q)$ is $s+2$.
If for certain $a$ between $a_*$ and $a_o$ the polyhedron
$P(q,a)$ were nonempty, then for any $a'$ between $a_*$ and $a$
it would have full dimension $s+1$, so that the word $q$ would belong
(by Lemma 1.5) to $\T(a')$ and also to $\T(a_o)$. Thus all the sections
$P(q,a)$ are empty for $a$ after $a_*$ and the maximum of $l$ is attained
on $P(q,a_*)$.

Inversely, if the maximum of $l$ is attained on $P(q,a_*)$, then
all the polyhedra $P(q,a)$ for $a$ after $a_*$ are empty. Further,
if the dimension of $P_I(q)$ is $s+2$, then some of the fibers
$P(q,a)$ have the dimension $s+1$ and thus all of them before $a_*$, as
$P(q,a_*)$ is nonempty.
\enddemo

To investigate implications of the criticality of a point
we introduce a graph and a flow on it. The vertices of the
graph correspond to $x$-variables in $W$ and the edges
to constraints defining the polyhedra $P(q)$.

Let $q$ be a word in $\A$ of length $n$. The graph $\Ga(q)$ has
$2(s+1)+n$ vertices
$v^-_0,\ldots,v^-_s; v_1,\ldots,v_n; v^+_0,\ldots,v^+_s$.
The (oriented) edges of $\Ga(q)$ are of two types:
first, independent of $q$, are following:
connecting each $v^-_j$, $j=0,\ldots,s$ to $v_1$;
$v_1$ to $v_2$, $v_2$ to $v_3, \ldots, v_{n-1}$ to $v_n$ and
$v_n$ to each of $v^+_j$, $j=0,\ldots,s$.
This  $q$-independent part is a tree
and will be denoted as $\Ga$; the edges of this tree will be
referred to as $m$-edges.
The edges of second type are $q$ specific and connect
$v^-_j$ to $v_{i^j_1}$; this latter one to $v_{i^j_2}$ and so on
until $v_{i^j_{\vert I_j\vert}}$. This last vertex is connected to $v^+_j$.
(Recall that $I_j=(i_1,\ldots,i_{\vert I_j \vert})$ is the subset of
$\{1,\ldots,n\}$ consisting of indices
$i$ such that $q_i=j$; if the subset is empty $v^-_j$ is connected
directly to $v^+_j$).
We will call them
$l$-edges and will mark them by the corresponding letters of $\A$.

The edges of the graph $\Ga(q)$ correspond to constraints defining
the polyhedron $P(q)$; $m$-edges
corresponding to inequalities defining the cone $C$ and
$l$-edges to equalities
defining $W(q)$.

Recall, that a {\it closed flow} in an oriented graph
is a function on its edges such
that the Kirchhoff's current law is satisfied: for any vertex
the sum of all flows (values of the
function) over the {\it in}-edges is equal to
that over all {\it out}-edges.

Recall also some basic facts from the linear programming theory
which will be used in the proof of the following Proposition
(see, e.g. [FF]):

\proclaim{Fact}
A. Let $P$ be a polyhedron in $\Real^S$ defined by a system of linear
equations and inequalities
$$
\psi_\alpha \geq 0; \phi_\beta =0;
$$
$\alpha = 1,\ldots,A; \beta = 1,\ldots,B; \phi$'s and $\psi$'s --- linear
inhomogeneous functions. Let the maximum of a linear function $l$ be attained
on a face $F$ of the polyhedron, and $y$ is a relatively
interior point of the face. Then there exists a linear
combination (whose coefficients are called Lagrange multipliers)
of the linear functions constraining the polyhedron (i.e. $\psi$'s
and $\phi$'s),
with {\it nonnegative} coefficients for $\psi$'s, whose sum with
$l$ is a constant. Moreover, the linear combination can be
chosen in such a way, that the coefficient for $\psi_\alpha$
is {\it positive} exactly when $\psi_\alpha(y)=0$ (LP duality).

B. Inversely, if a linear combination
of constraining function with non positive coefficients for the linear
functions entering inequalities defining $P$ plus
$l$ is constant, and all the constraining functions with nonzero
coefficients vanish at $y \in P$, then
$l$ achieves
its maximum over $P$ at $y$.
\endproclaim

\proclaim{Proposition 2.2} Let $a_*$ be critical
and the word $q$ either vanishes
from thesaurus or appears there when $a$ varies through $I$. Then there
exists a non-zero closed flow on $\Ga(q)$, and
if $x$ is a relatively interior point of $P(q,a_*)$, then
$m$-edges in the support of the flow correspond exactly to
those inequalities which becomes equalities on $x$ and the flow
through any of these $m$-edges is positive.
\endproclaim

\demo {Proof} Choose Lagrange multipliers
$$
\eqalign{
m^-_j \geq 0\ \for \ & x_1-x^-_j \geq 0 \cr
m_i \geq 0 \ \for \ & x_{i+1}-x_i \geq 0 \cr
m^+_j \geq 0\ \for \ & x^+_j-x_n \geq 0
}
\tag2.3
$$
for inequalities defining cone $C$.

The constraints, defining the linear subspace $W(q)$ are either
$$
x_i - x^-_j -a_j=0, \or x_i-x_{i'}-a_j=0, \or x^+_j-x_i-a_j=0,
$$
and we attach Lagrange multipliers $l^-_j, l_{i,i'}$ and $l^+_j$ to
them respectively. Thus to each constraint and, consequently,
to each edge of $\Ga(q)$ a Lagrange multiplier is associated.

One more set of multipliers we will need for the constraints
forcing $a$ to belong to the line through $a_i, a_o$.
(Notice that there is no need to introduce multipliers for $a_j \geq 0$
as the inverse velocities are positive by assumption.)

So, according to the Fact of the
linear programming theory (as stated above),
if $a_*$ is critical, and $(x,a_*)$ is an interior point
of $P(q,a_*)$, we can choose a set of coefficients (Lagrange
multipliers), such that the resulting linear combination plus $l$ is constant.

Consider now these multipliers as defining a flow $\Phi$ on $\Ga(q)$.
Indeed, to each of them corresponds a unique edge in the graph,
so that the $x$-part of the corresponding function is the difference
of $x$'s at the ends of the edge. The equality of the linear combination
to $-l$ plus a constant
means exactly that thus defined flow is closed: the coefficient for any
of $x$'s is the algebraic sum of flows to the corresponding
vertex, and $l$ does not depend on $x$'s (in other words,
the Lagrange multipliers define a closed 1-cocycle on $\Gamma (q)$).

The statement about $m$-edges in the support of the flow is
tantamount to the LP duality.
\enddemo

A cycle in a graph is called simple if passes through any edge at most once.
A simple (non oriented) cycle in the oriented graph $\Ga(q)$
is said to be {\it subordinated to the flow} $\Phi$,
if it contains $m$-edges only from the support of $\Phi$ and
traverses those $m$-edge
in accordance with their orientation. To any simple cycle a closed
flow corresponds;
it takes a constant (positive) value on all edges of the cycle.
Such closed flows will be also called simple and subordinated if the
underlying cycle is.

Simple subordinated cycles are important as they
generate bounded resonances at $a_*$. A {\it resonance} is a vanishing
linear integer combination of
$a_j$'s. A resonance will be called bounded if
the sum of absolute values of its coefficients does
not exceed the number of edges in $\Ga(q)$.

\proclaim{Lemma 2.4} Let $a_*$ be critical; $x$ be an interior point in
$P(q,a_*)$ and $\Phi$ be a closed flow defined by the Lagrange
multipliers at $x$. Then to any simple subordinated to $\Phi$ cycle
a bounded resonance at $a_*$ corresponds.
\endproclaim

\demo {Proof} Any edge of the cycle corresponds either to
equality $x_i-x_{i'}-a_j=0$, or to equality $x^\cdot_\cdot -x^\cdot_\cdot=0$
(the $m$-edges are in the cycle if the corresponding Lagrange
multiplier is positive only, what makes them to be equalities).
Summing up all of them we obtain a bounded resonance at $a_*$.
\enddemo

\proclaim{Lemma 2.5} For any closed flow $\Phi$ subordinated cycles exist.
Moreover, any closed flow can be decomposed into a positive
linear combinations of simple subordinated flows.
\endproclaim

\demo{Proof} Reverse the orientations of all $l$-edges with
negative flow, so that the flow through any edge is nonnegative.
A subordinated cycle can be then found by the following
algorithm: pick any edge in the support of the flow and go
along the arrow.
In the reached vertex choose new adjoining edge
along which the movement according to its orientation is
possible --- such an edge always exists due to the closeness of the flow.
Iterating we will reach a vertex already seen at a stage, thus getting
a subordinate cycle. To prove the decomposition part of the Lemma,
define the closed flow on this chosen cycle by assigning to each
edge in it the minimal flow of the edges gone
through. Subtracting the resulting closed flow from the initial one
we will obtain a flow with a smaller support. Iteratings finish the
prove.
\enddemo

\head 3. Simple Resonances \endhead

We say that that the critical point $a_*$ is
simple if all bounded resonances at the point
are integer multiples of a single resonance $n \cdot a_*=0$.
The condition of simplicity
restricts strongly the structure of possible simple cycles
subordinated to the flows associated with the critical point $a_*$.
In fact one can prove, that such a the support of such a flow is
necessarily a simple
cycle. We will prove here only a weaker result needed in what follows.

\proclaim{Proposition 3.1} Let $q \in \T(a_i) -\T(a_o)$;
$a_* \in I$ be a simple critical point on the segment $I$; $x$
be an interior point of  $P(q, a_*)$
and $\Phi$ be the associated closed flow.
Then at most one of the edges $x^-_j \to x_1$
and at most one of the edges $x_n \to x^+_j$ belong to the support
of $\Phi$.
\endproclaim

\demo{Proof}
Consider a cycle subordinated to the flow $\Phi$. For any mark $\bold j$
we define a $\bold j$-segment in the cycle as a sequence of $l$-edges
with this mark bounded by edges of other (necessarily not $l$-) types.

First, we will prove, that for any simple cycle subordinated
to the associated flow and for any mark $\bold j$ there is at most one
$\bold j$-segment in the cycle.
Indeed, let $s_1, s_2$ be two $\bold j$-segments separated by
some pieces of the cycle $c_1, c_2$, so that the whole cycle has the form
$$
-s_1-c_1-s_2-c_2-s_1-.
$$
We can now form two new cycles joining the ends of $c_1$ and $c_2$ by
$\bold j$-segments. These two cycles are clearly simple and subordinate
as the $m$ edges remain unchanged.

An easy check shows that the closed  flow
whose support is initial simple cycle is now the sum of thus constructed
closed flows. Decomposing if necessary these flows further we arrive at a
stage to the situation when each of the new cycles has at most one $\bold j$
segment. If the
initial cycle had more than one $\bold j$ segment, then among these cycles
exist both cycles going through $\bold j$ edges in positive and negative
directions.

Each of the cycles generates a nontrivial (as $n_j\neq 0$)
resonance, a multiple of $n$ by assumption.
The fact that both positive and negative
multiples of $n$ occur means that both $l$ and $-l$
can be represented as linear combinations of the constraining functions
with nonnegative $m$'s (as the constructed
flows are subordinated), and thus both
$l$ and $-l$ achieve their maxima on $P_q$ at the point $(x,a_*)$ (part B of
the linear programming Fact).
It follows that $P_I(q)=P(q,a_*)$ in contradiction with the assumption,
that $P_I(q)$ has dimension $s+2$.

Thus each of
the simple subordinated cycles has at most one $\bold j$ segment,
and the direction of traversing them is the same for all cycles.

Assume now that the support of the associated flow contains two ($m$-)
edges $x^-_{j_1} \to x_1$ and $x^-_{j_2} \to x_1$. Let $\gamma_1,\gamma_2$
be two simple subordinated cycles containing these edges correspondingly.
Let $s$ be the segment of the cycle $\gamma_2$ starting in $x_1$  and ending
at the beginning of the first ${\bold j_1}$ edge. Then cutting short along
$\bold {j_1}$-edges to $x^-_{j_1}$ and from there to $x_1$ we obtain
a new subordinate cycle. This cycle contains no ${\bold j_2}$
edges as they obviously cannot belong to $s$, but contain more
${\bold j_1}$ edges than $\gamma_2$. Thus the resonance defined by the
cycle cannot be a multiple of $n$ --- a contradiction. Similar
reasoning gives $x^+_\cdot$ part.
\enddemo

\proclaim{ Corollary 3.2} For any interior point $x$ of $P(q,a_*)$
at most one of inequalities $x^-_j \leq x_1$ ($x_n \leq x^+_j$)
becomes equality.
\endproclaim

\demo{Proof} It follows immediately from Propositions 2.2 and 3.1.
\enddemo

\head 4. In-Out Correspondence \endhead

Now we will prove the crucial result: if the critical point
on the segment $I$ is simple (that is satisfies at most one
resonance modulo natural multiples) then the number of words
disappearing from the thesaurus equals the number of words
appearing there. To do it we construct the {\it in-out correspondence}
as follows.

Let $x$ be a vector in $W_x$, and $q$ a word of length $n$. They
together define a marking of $\Real$, that is a function
from reals to the subsets of $\A$: one marks $x^\pm_j$ and $x_i$
with $q(i)=j$ by the letter ${\bold j}$ and assembles multiple letters
at each point.
We will call a marking {\it good in the middle} if there is an
interval, such that total of $(s+1)$ letters are on the left side,
same number on the right side of the interval and the remaining $n$ markings
are simple, that is each point in the interval is marked by
at most one letter. A good in the middle marking defines
a word of length $n$: one just reads these middle $n$ letters
in their natural order. This word we denote by $Q(x;q)$; it
coincides with $q$ if $x$ is an
interior point of $C$.

Let $a_*$ be our simple critical point, the word $q$ belongs to
$\T(a_i)-\T(a_o)$ and $P(q,a_*)$ be the face of the polyhedron
$P_I(q)$ where $l$ attains its maximum. Let $x_*$ be a relatively
interior point of the polyhedron. Choose a line $L$ through
$(x_*,a_*)$ which projects onto $I$ under $p_a$ and such that
one of the halflines on which $(x_*,a_*)$ divides it belongs
to $P_I(q)$. We will call this halfline {\it the $a_i$-side}, and
the other one --- {\it the $a_o$-side}.

When a line in $W$ is given, such that it projects onto $I$, we
can consider $l$ as a parameter on the line. In this case the
coordinates $x^-_j, x_i, x^+_j$ become linear function of $l$ and
a movement along the line can be considered as an time
evolution of these points steady moving in $\Real$. We will be using this
convenient terminology throughout this section.

\proclaim{Lemma 4.1} Points of the line $L$ close to $(x_*,a_*)$ enough
define a good in the middle marking of $\Real$.
\endproclaim

\demo{Proof} The statement is trivial for points on the $a_i$-side
of the line. If that is not the case for $a_o$-side, then
a pair of the marked points coincide identically along the line.
The pair cannot include any of $x_i$ points, as each of them is separated
from the rest on the $a_i$ side. Equally, it cannot be a pair
$x^-_{j_1},x^+_{j_2}$. So, assume that $x^-_{j_1}=x^-_{j_2}$
identically on the line. To affect the middle part of the marking
on the $a_o$-side,
the pair should move through $x_1$ when $a=a_*$. That yields
$x^-_{j_1}=x^-_{j_2}=x_1$ at $x$ -- a relatively interior point
of $P(q,a)$, and thus
$$
x_1 - x^-_{j_1}=0; x_1-x^-_{j_2}=0
$$
at $x_* \in P(q,a_*)$.
This contradicts Corollary 3.2

The $x^+_\cdot$ case is similar.
\enddemo

Thus we have a good in the middle marking defined by $(x_*,a_*)$
on the $a_o$ side of the line and form the word $q'=Q(q,x)$.
This correspondence $q \mapsto q'$ is called {\it in-out}
one.

To use it we have to prove first that it is unambiguous.

\proclaim{ Proposition 4.2} The in-out correspondence
is defined unambiguously, that is does not depend on the choice
of the interior point $x_* \in P(q,a_*)$ and of line through $x_*$.
\endproclaim

\demo{Proof} Let $L,L'$ be two lines through a relatively interior point
$x$ of $P(q,a)$ which $p_a$ maps onto $I$, such that $a_i$-sides
of both of them belong to $P_I(q)$. Assume that their $a_o$-sides
define {\it different} words $q',q''$.
Let $V$ be the two-dimensional plane spanned by $L,L'$.
This plane is fibered by the level lines of $p_a$ and both the
lines $L,L'$ are transversal to the fibration. These lines cut a segment
in each fiber of $p_a$. The difference of the words defined by the
middle parts of the corresponding markings implies that inside the
segment a couple of these middle points coincide. The finiteness
of the number of such couples yields
existence of a line in $V$ between $L$ and $L'$ which is projected
onto $I$ and along which a pair of $x$-coordinates coincide identically.
The pair cannot contain any of $x_i$'s, as it would contradict the
assumption that $a_i$ parts of both $L,L'$ are in $P_I(q)$, or
be $x^-_{j_1},x^+_{j2}$. Thus it is a pair $x^-_{j_1},x^-_{j2}$, and
at $x_*$ he have
$$
x_1 - x^-_{j_1}=0; x_1-x^-_{j_2}=0,
$$
in contradiction with Corollary 3.2.

Let now $x,x'$ be two different relatively interior points in $P(q,a)$.
Choose two lines $L \ni x,L'\ni x'$ laying in a two-dimensional
plane $V$, projecting onto $I$ with their $a_i$ parts in $P_I(q)$.
The assumption, that the words defined by their $a_o$ parts are different
means again, that between $L$ and $L'$ in $V$ there is a line along
which some pair of $x$ coordinates coincide. Reasonings as above prove it
impossible.
\enddemo

Further thing we need to show is that if $q$ leaves thesaurus
when moving from $a_i$ to $a_o$, then $q'$ appears there.

\proclaim{Proposition 4.3} If $q,q'$ are in in-out correspondence, and
$q \in \T(a_i)-\T(a_o)$, then $q' \in \T(a_o)-\T(a_i)$.
\endproclaim
\demo{Proof} By the construction of $q'$, the polyhedron $P(q')$
has full dimension - the fact that the marking defined
by a point $(x',a')$ on the $a_o$-side of the line $L$ passing
through $(x_*,a_*)$ is good in the middle means that
 all inequalities constraining $P(q')$ are strict
at the point corresponding to the marking.
So we
know that $q' \in \T(a_o)$ and we need to prove only that it
does not belong to $\T(a_i)$. Assume it does.
That means, that the minimal value of $l$ on $P(q')$ is not $l(a_*)$.
Then, as
$P(q')$ is a convex polyhedron, there exists a segment $\tilde I \subset W$
having $(x',a')$
as an endpoint and
such, that the minimal value of $l$ on it is less than that on $a_*$,
whose points define good in the middle markings.
Let two-dimensional
plane $V$ be spanned by this segment and the point $(x_*,a_*)$.
Apparently, $V$ contains $L$.
Choose a point $(x,a)$ on the $a_i$-side of $L$ close enough to $(x_*,a_*)$
and take a line $L'$ passing
through this point parallel to $\tilde I$. As $l$ is bounded
by $l(a_*)$ on the segment $P(q)\cap L'$, there is an
inequality among those defining $C$, which becomes equality at a point
$(x'',a'')$
of $L'$ and
at $(x_*,a_*)$ also. The line $L''$ passing through the points
$(x'',a''), (x_*,a_*)$ intersects $\tilde I$ in a point where
the said inequality is equality, but which defines a good in the
middle marking (corresponding to an interior point of $P(q')$).

If the aforementioned
equality is of the form $x_i=x_{i+1}$, then to leave
the ordering of
the markings corresponding to the word $q'$ in the course of movement
along $\tilde I$ unaffected,
these two points of $\Real$ should
leave the set of the middle $n$ points during the movement,
which is possible only when two points $x^-_j, x^-_{j'}$
or
$x^+_j, x^+_{j'}$ join the set of middle points.
To do it they have to exchange with $x_1$ ($x_n$, respectively) at
$(x_*,a_*)$, which again contradicts Corollary 3.2.

If the equality is of the form $x_1=x^-_j$, then to leave
the middle markings defining $q'$ unaffected,
$x_1$ should transpose with some
$x^-_{j'}$, which is to join the middle $n$ markings defining $q'$
during the movement along $I$, and $j \neq j'$, as $x_j$ cannot be
among the middle $n$ markings defining $q'$. Thus at $(x_*,a_*)$
$x_1=x^-_j=x^-_{j'}$ contradicting  Corollary 3.2 again.
The same reasonings prove that $x_n=x^+_j$ is either impossible. Thus
our assumption was wrong and $q' \not\in \T(a_i)$
\enddemo

The last step is to prove the reflectivity of the in-out
correspondence.

\proclaim{Proposition 4.4} If $q \mapsto q'$, then $q' \mapsto q$.
\endproclaim
\demo{Proof}
A marking of $\Real$ will be called $q$-consistent, if there
exists its small deformation resulting in a good in the middle marking
corresponding to the word $q$. If $q \mapsto q'$ and $(x_*,a_*)$
is the point used to define this in-out correspondence, then
the marking defined by $x_*$ is, obviously, both $q$- and $q'$-consistent:
moving along the line $L$ one of whose half-lines is pointed inwards $P(q)$
we will get both words.

Let $L'$ be a line in the fiber $p_a^{-1}$ passing through the point
$(x_*,a_*)$, such that points of one of half-lines close to $(x_*,a_*)$
define $q'$-consistent marking. Then the points of the other half-line
close to $(x_*,a_*)$ define a $q$-consistent marking.
Indeed, if it were not the case, then one of the inequalities
defining $P_I(q)$ would be both strictly violated and strictly
satisfied by different good in the middle markings defining $q'$ and
close to $x_*$. Thus on a line in the two-dimensional plane spanned by
$L$ and $L'$ passing through $(x_*,a_*)$ this inequality becomes
an equality.

Now the usual case analysis applies: if the inequality were of the form
$x_i \leq x_{i+1}$, then to have the word $q'$ unaffected by the
transposition of the points, these two had to leave the set of
middle $n$ markings under movement along $I$, that is some two
of $x^-_j$'s or some two of $x^+_j$ had to join the middle set, which is
impossible. Equally impossible is the option $x^-_j=x_1$ (here $x_1$ had
to leave the middle set exchanging with some $x_-^{j'}$ underway), and
$x^+_j=x_n$ by the symmetric reasons.

Thus, if the point moves along $L'$ passing through $x_*$, the
markings corresponding to the point move steadily in $\Real$, and under the
passage of  $x_*$, the $q'$-consistent
marking becomes a $q$-consistent one.
Let $(\tilde x_*, a_*)$ be the point in $P(q',a_*)$ corresponding to
the point $(x_*,a_*)$ in $P(q,a_*)$. The last statement says
that the tangent cone to $P(q',a_*)$ at $(\tilde x_*, a_*)$
is of dimension at most the dimension of the tangent cone to $P(q,a_*)$
at $(x_*,a_*)$. On the other hand, the Proposition 4.2 implies that
one can identify the polyhedron $P(q,a_*)$ with a face of $P(q',a_*)$.
That means that these two polyhedra coincide and that the point
$(\tilde x_*, a_*)$ is interior in $P(q',a_*)$. The movement
of a point along $L$ defines a deformation of the marking of $\Real$;
taking its `after $a_*$ part we get a halfline passing through
$(\tilde x_*, a_*)$ whose point close to the origin are interior
in $P(q',a_*)$.
That, apparently, yields $q' \mapsto q$ and thus the reflectivity of the
in-out correspondence.
\enddemo

\proclaim{Corollary 4.5} The size of thesaurus is constant in two
endpoints $a_i$, $a_o$ of a segment containing a single simple
resonance point $a_*$.
\endproclaim

\head 5. Proof of the Main Theorem \endhead

\proclaim{ Theorem} The size of thesaurus $\#\T(a)$ for rationally
linearly independent inverse velocities $a_j$ is independent of
velocities and is given by the formula
$$
\sum_{k=0}^{\min{n,s}}{k !}{s \choose k} {n \choose k}.
$$
\endproclaim

\demo{Proof} Let $a_1, a_2$ be inverse velocities.
Choose a piecewise linear path joining them in the space of
inverse velocities, such that endpoints of each segment of the
path were linearly independent over $\Rat$ and each point of the
path belonged to at most one bounded resonance hyperplane
$(n,\cdot)=0$. Such a path can be chosen as the number of bounded
resonances is finite and the points where multiple resonances
occur have codimension 2.
Using the Corollary 4.5 we see, that
the size of the thesaurus is constant in the endpoints of
the segments forming the path and thus the sizes of thesaurus
in $a_1$ and $a_2$ are equal. To find it we choose a special
vector of inverse velocities. Namely, let $a_0=1$ and $a_1,\ldots,a_s$
be arbitrary numbers larger than $n$, such that the whole
tuple $a_0,\ldots,a_s$ is $\Rat$-independent. Then
the thesaurus can be easily described: a word $q$ belongs to the
thesaurus exactly when it contains at most one letter $\bold j$
with $j \geq 1$. The number of such words can be calculated immediately.
Indeed, each of the words is specified by the number $k$ of letters
${\bold j}$ with $j \geq 1$; by the
positions in the word
of letters of these letters (${n \choose k}$ possibilities); by the set of
letters ${\bold j}$ with $j \geq 1$ used (${s \choose k}$ possibilities) and
by one of $k!$ variants of their allocation there. Summing
it all up one gets the stated answer.
\enddemo

\head 6. Concluding Remarks \endhead

\subhead (6.1) \endsubhead
As a corollary of the presented result we get the remarkable
symmetry noted in [AMST]: the size of $n$-thesaurus in $\Real ^{s+1}$
is a function symmetric in $n$ and $s$. Actually, as it was shown
in [AMST] this property almost characterizes this function and thus
it would be very interesting to have a construction directly
proving this symmetry. I believe, that the size of $n$-thesaurus
can be described in terms of the facet combinatorics of an appropriately
chosen polyhedron and that $n \leftrightarrow s $ symmetry would
follow from Dehn-Sommerville relations. Such approach would give
a much more clear insight into the combinatorics of symbolic
trajectories, but so far I do not know, how to realize it.

\subhead (6.2) \endsubhead
The Lemma 1.5 provides in principle an algorithmic method to define
whether a word is a piece of symbolic trajectory of a billiard.
One has to check to compatibility of a system of linear inequalities,
which can be effectively done.

\subhead (6.3) \endsubhead
Using the geometric approach of [AMST] one can generate a subdivision
of the $s$-dimensional torus parameterizing the billiard trajectories
with given velocity
into the convex polyhedra corresponding to different words of the
thesaurus of a size $n$. The volumes of these polyhedra closely relate
to the reccurence function of symbolic trajectories, that is the
size of word in which all words of the $n$-thesaurus appear. The moving
of inverse velocity through a resonance results in the degeneration of
some of these polyhedra, and thus enables one to describe the asymptotics
of the minimal of their volumes. That gives an approach to the investigation
of the reccurence function. I hope to return to the question in a separate
paper.

\heading {\bf REFERENCES} \endheading

\item{[AMST]} Arnoux, P., Mauduit, C., Shiokawa, I., Tamura, J.-I.,
Complexity of sequences defined by billiard in the cube.
{\it Bull. Soc. math. France, {\bf 122}, 1-12 (1994)}.

\item{[B]}
Bruckstein, A. M., Self-similarity properties of digitized straight lines.
{\it In:
Vision geometry, Proc. AMS Spec. Sess., 851st Meet., Hoboken/NJ (USA)
    1989, Contemp. Math. 119, 1-20 (1991).}

\item{[FF]}
Ford, L.R., Fulkerson, D.R. {\it Flows in Networks}, Princeton, 1962.

\item{[LP]}Lunnon, W. F., Pleasants, P. A. B,
Characterization of two-distance sequences.
{\it J. Aust. Math. Soc., Ser. A {\bf 53}, No. 2, 198-218 (1992).}

\item{[MH]} Morse, M., Hedlund, G.A. Symbolic dynamics II. Sturmian
trajectories. {\it Amer. J. Math. {\bf 62}, 1-42 (1940).}

\item{[R]} Rauzy, G. Mots infinis et arithmetique, {\it In:
Automata on Infinite Words, Lect. Notes in Comp. Sciences, {\bf 192},
165-171}.

\item{[S]} Stolarsky, K. B., Beatty sequences, continuous
fractions and certain shift operators, {\it
Can. Math. Bull. {\bf 19}, 473--482 (1976)}.

\end